\documentclass[aps,twocolumn,preprintnumbers,showpacs,showkeys,nofootinbib%
]{revtex4}
\usepackage{epsfig}
\usepackage{amssymb,amsmath,amsfonts,amsthm,graphicx,psfrag}
\graphicspath{{./Figures/}}                 
\newcommand{\noi}{\noindent}
\newcommand{\beq}{\begin{equation}}
\newcommand{\eeq}{\end{equation}}
\newcommand{\bea}{\begin{eqnarray}}
\newcommand{\eea}{\end{eqnarray}}

\newcommand{\Tab}[1]{Table~\ref{#1}}

\newcommand{\tr}{\operatorname{Tr}}

\begin{document}

\title{Study of the thermal abelian monopoles with proper gauge fixing}

\author{V.~G.~Bornyakov}
\affiliation{High Energy Physics Institute, 142280 Protvino, Russia \\
and Institute of Theoretical and Experimental Physics, 117259 Moscow, Russia}

\author{V.~V.~Braguta}
\affiliation{High Energy Physics Institute, 142280 Protvino, Russia}

\begin{abstract}

The properties of the thermal abelian monopoles are studied in the deconfinement phase of
the $SU(2)$ gluodynamics. To remove effects of Gribov copies the simulated annealing
algorithm is applied to fix the maximally abelian gauge. Computing the density of the
thermal abelian monopoles in the temperature range between $1.5~T_c$ and $6.9~T_c$ we show, by comparison with earlier results, that
the Gribov copies effects might be as high as 20\% making  proper gauge fixing mandatory.
We find that in the infinite temperature limit the monopole density converges to its value in
3-dimensional theory. To study the interaction between monopoles we calculate the monopole-monopole and monopole-antimonopole correlators
at different temperatures in the region $(1.5~T_c, 6.9~T_c)$. Using the result of this 
study we determine the screening mass, monopole-monopole coupling constant, monopole size and 
monopole mass. In addition we  check the continuum limit of 
our results. 
\end{abstract}

\keywords{Lattice gauge theory, deconfinement phase, thermal monopoles,
Gribov problem, simulated annealing}

\pacs{11.15.Ha, 12.38.Gc, 12.38.Aw}

\maketitle

\section{Introduction}

Our study of the thermal monopoles is motivated by the hypotheses that some of the quark-gluon plasma
properties may be dominated by a magnetic component~\cite{Liao:2006ry,Chernodub:2006gu,Shuryak:2008eq}, at least for the range of the temperature values close to the transition.
The monopoles or magnetic  vortices might be responsible for the very low viscosity indicating that the quark-gluon plasma is an ideal liquid.

The way one can study the monopoles properties in lattice nonabelian gauge theories is via an abelian projection after fixing the maximally
abelian gauge (MAG) \cite{'tHooft:1981ht,'tHooft:1982ns}. Since first lattice 
studies of this gauge \cite{Kronfeld:1987vd,Suzuki:1989gp} the properties of 
the abelian monopole
clusters had been investigated in a numerous papers both at zero and nonzero
temperature. The evidence was found that the nonperturbative properties of the gluodynamics
such as confinement, deconfining transition, chiral symmetry breaking, etc.  are closely related to
the abelian monopoles defined in MAG \cite{Suzuki:1992rw,Chernodub:1997ay,Ripka:2003vv}. This was called a monopole dominance.

It was shown in Ref.~\cite{Chernodub:2006gu} that thermal monopoles in Minkowski
space are associated with Euclidean monopole trajectories wrapped around the temperature direction of
the Euclidean volume. So the density of the monopoles in the Minkowski space is given by the average
of the absolute value of the monopole wrapping number.
First numerical investigations of such wrapping trajectories were performed in Refs.~\cite{Bornyakov:1991se} and~\cite{Ejiri:1995gd}.
A more systematic study of the thermal monopoles in $SU(2)$ Yang-Mills
theory at high temperature has been performed in Ref.~\cite{D'Alessandro:2007su}. It was found
in ~\cite{D'Alessandro:2007su} that the density of thermal monopoles $\rho$
was independent of the lattice spacing as it should be for a physical observable.
It was concluded that the monopole density was well described by 
$\rho \propto T^3/(\log T/\Lambda_{eff})^2$ with
$\Lambda_{eff} \sim 100$ MeV while the behaviour $\rho \propto T^3/(\log T/\Lambda_{eff})^3$,
predicted by dimensional reduction arguments, was compatible with
data for  $T > 5\ T_c$.

The density--density spatial correlation functions has been also computed in ~\cite{D'Alessandro:2007su}.
It has been shown that there is
a repulsive (attractive) interaction for a monopole--monopole (monopole--antimonopole) pair,
which at large distances might be described by a screened Coulomb potential with  a screening length of
the order of 0.1 fm.

It is known that the Gribov copies effects are strong in the MAG ~\cite{Bali:1996dm}.
In ~\cite{Bali:1996dm} a conclusion has been made that results for gauge noninvariant observables can be
substantially corrupted by inadequate gauge fixing. For nonzero temperature the effects of Gribov copies
were not investigated so far.
In this paper we would like to close this gap. Following ~\cite{Bali:1996dm} we apply the simulated annealing algorithm with 10 gauge copies for every configuration to solve the problem of Gribov copies by approaching the global maximum of the gauge fixing functional. We present our results for the density of the thermal monopoles and the correlation functions and compare them with the results of Ref.~\cite{D'Alessandro:2007su}.
We show that the Gribov copies effects are indeed large. We also present results for screening mass 
and coupling constant of monopoles interaction, the size and mass of monopole.

The paper is organized as follows. In the next section we briefly review 
the details of the simulation. In the section III we study the dependence of 
monopole density on the temperature. In section IV the monopole-monopole and 
monopole-antimonopole correlators are calculated at different temperatures.  
Using the result of this calculation we determine the size of monopole, 
Debye screening mass, monopole-monopole coupling constant. In section V 
we calculate the dependence of the monopole mass on the temperature. 
In section VI we consider the continuum limit of our results. In last section 
the results of this paper are summarized.

\section{Simulation details}
We study the SU(2) lattice gauge theory with the standard Wilson action

\beq
S  = \beta \sum_x\sum_{\mu >\nu}
\left[ 1 -\frac{1}{2}\tr \Bigl(U_{x\mu}U_{x+\mu;\nu}
U_{x+\nu;\mu}^{\dagger}U_{x\nu}^{\dagger} \Bigr)\right], \nonumber
\label{eq:action}
\eeq

\noi where $\beta = 4/g_0^2$ and $g_0$ is a bare coupling constant. The
link variables $U_{x\mu} \in SU(2)$ transform  under gauge
transformations $g_x$ as follows:

\beq
U_{x\mu} \stackrel{g}{\mapsto} U_{x\mu}^{g}
= g_x^{\dagger} U_{x\mu} g_{x+\mu} \; ;
\qquad g_x \in SU(2) \,.
\label{eq:gaugetrafo}
\eeq


\noi Our calculations were performed on the asymmetric lattices with
lattice volume $V=L_t L_s^3$, where $L_{t,s}$ is the number of sites in
the time (space) direction. The temperature $T$ is given by

\beq
T = \frac{1}{aL_t}~,
\eeq

\noi where $a$ is the lattice spacing.

The MAG gauge is fixed by finding an extremum of the gauge functional

\beq
F_U(g) = ~\frac{1}{4V}\sum_{x\mu}~\frac{1}{2}~\tr~\biggl( U^{g}_{x\mu}\sigma_3 U^{g\dagger}_{x\mu}\sigma_3 \biggr) \;,
\label{eq:gaugefunctional}
\eeq

\noi with respect to gauge transformations $g_x$. We apply the simulated annealing (SA) algorithm which proved to be very efficient for this
gauge \cite{Bali:1996dm} as well as for other gauges such as center gauges and Landau gauge. To further decrease the Gribov copy effects we generated 10 Gribov copies starting every time gauge fixing procedure from a randomly selected gauge copy of the original Monte Carlo configuration.

In \Tab{tab:statistics} we provide the information about the gauge field ensembles used in our study.


\begin{table}[ht]
\begin{center}
\begin{tabular}{|c|c|c|c|c|c|} \hline
 $\beta$ & $a$[fm] & $L_t$ & $~L_s~$ & $T/T_c$ &
$N_{meas}$ \\ \hline\hline
  2.43   & 0.108 & 4  & 32  & 1.5     & 1000    \\
  2.5115  & 0.081 & 4  & 28  & 2.0     & 400      \\
  2.635   & 0.054 & 4  & 36  & 3.0     & 500      \\
  2.635   & 0.054 & 8  & 48 & 1.5 & 1000 \\ 
  2.70     & 0.046 & 4  & 36  & 3.6     & 200      \\
  2.74     & 0.041 & 4  & 36  & 4.0     & 100      \\
  2.80     & 0.034 & 4  & 48  & 4.8     & 1000      \\
  2.85    & 0.029 & 4  & 48  & 5.7     & 100       \\
  2.90    & 0.025 & 4  & 48  & 6.3    & 100         \\
  2.93    & 0.023 & 4  & 48  & 6.9     & 1000       \\ \hline

\end{tabular}
\end{center}
\caption{Values of $\beta$, lattice sizes, temperatures, number of
measurements and number of gauge copies used throughout this paper.
To fix the scale we take $\sqrt{\sigma}=440$ MeV.
}
\label{tab:statistics}
\medskip \noindent
\end{table}


\section{Thermal monopole density}

We locate the wrapped monopole trajectories using a standard prescription. The monopole current is
defined on the links $\{ x, \mu \}^*$ of the dual lattice and take integer values $j_{\mu}(x)=0, \pm 1, \pm 2$.
The monopole currents form closed loops combined into clusters. Wrapped clusters are closed through the lattice boundary.
For a given cluster the wrapping number $N_{wr}$ is defined as
\beq
N_{wr} = \frac{1}{L_t} \sum_{j_4(x) \in cluster} j_4(x)
\eeq
It takes values $0, \pm 1, \pm 2 ...$.

\begin{figure}[tb]
\centering
\includegraphics[width=6.0cm,angle=270]{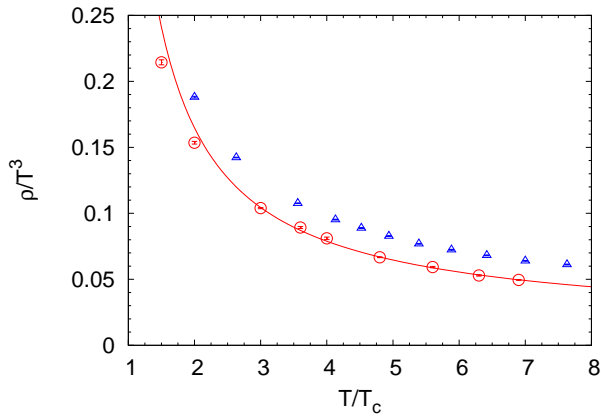}
\caption{Thermal monopole density (circles) as function of temperature. For comparison data from Ref.~\cite{D'Alessandro:2007su}
are shown (squares). The curve shows fit to eq.(\ref{dimreduction}).
}
\label{fig:density}
\end{figure}

The density $\rho$ of the thermal monopoles is defined as
\beq
\rho = \frac{\langle~ \sum_{clusters}|N_{wr}| ~\rangle }{L_s ^3 a^3}\,.
\eeq
In Figure~\ref{fig:density} we show our data for the set of lattices listed in the Table~\ref{tab:statistics}.
The data of Ref.~\cite{D'Alessandro:2007su} are also shown for comparison. One can see strong Gribov copies effects from
the Figure~\ref{fig:density}. They are about 20\% at $T/T_c=2$ and increase to almost 30\% for $T/T_c=6.9$.
The dimensional reduction suggests for the density $\rho$ the following temperature dependence at high enough  temperature
\beq
\rho (T)= (c_\rho g^2(T) T)^3
\label{dimreduction}
\eeq
where the temperature dependent running coupling $g^2(T)$ is described at high temperature by the two-loop expression
with the scale parameter $\Lambda_T$:
\beq
g^{-2}(T) = \frac{11}{12\pi^2} \ln (T/\Lambda_T) + \frac{17}{44\pi^2}(\ln(2\ln (T/\Lambda_T))
\label{coupling}
\eeq
In  Ref.~\cite{D'Alessandro:2007su} the data for density $\rho$ were fitted to a fit function
\beq
\frac{A}{(\log(T/\Lambda_{eff}))^\alpha},
\eeq
which is motivated by a one loop expression for  $g^2(T)$.
A good fit was obtained for $T/T_c > 2$ with $\alpha=2$ and it was noted that $\alpha=3$, which corresponds to the one loop expression
for  $g^2(T)$ is compatible with data for  $T > 5\ T_c$.

\begin{figure}[tb]
\centering
\includegraphics[width=6.0cm,angle=270]{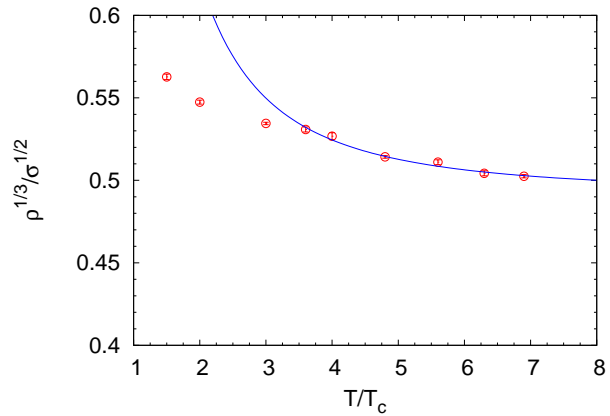}
\caption{The ratio $\rho^{1/3}(T)/\sigma_s^{1/2}(T)$  as a function of temperature. }
\label{fig:ratio}
\end{figure}

We fit our data to function (\ref{dimreduction}),(\ref{coupling}) with fitting parameters  $c_\rho$ and $\Lambda_T$. The reasonably good
fit with $\chi^2/dof = 1.4$ was obtained for $T \ge 3T_c$ with values for fit parameters $c_\rho=0.79(1)$, $T_c/\Lambda_T=2.52(6)$.
Thus behaviour of the density $\rho(T)$ is well described by the fitting function motivated by dimensional reduction prediction.
However, for the spatial string tension $\sigma_s(T)$ the parameter $\Lambda_T$ was found very different \cite{Bali:1993tz}:
$T_c/\Lambda_T=13.16(17)$. This implies that the dimensionless ratio $\rho^{1/3}(T)/\sigma_s^{1/2}(T)$ is decreasing with
temperature. \footnote{(Note that for  magnetic  screening mass $T_c/\Lambda_T=5.1(9)$ was found in
\cite{Heller:1997nqa}).}
This decreasing can be indeed seen from Figure~\ref{fig:ratio} where this ratio  is depicted. For  $ \sigma_s^{1/2}(T)$ we took the fitting
function obtained in \cite{Bali:1993tz}:
\beq
\sqrt{\sigma_s(T)} = c_s g^2(T) T
\label{sigma}
\eeq
with $c_s=0.369(14)$ and $T_c/\Lambda_T=13.16(17)$.

Note that decreasing seen in  Figure~\ref{fig:ratio} is slow, the ratio changes only about 10\% over our range of temperatures.
We fitted the ratio $\rho^{1/3}(T)/\sigma_s^{1/2}(T)$ to a polynomial fit
\begin{equation}
 \frac{\rho^{1/3}(T)}{\sigma_s^{1/2}(T)} = R+R_2/x^2\,,\,\, x=T/T_c\,,
\end{equation}
for $T/T_c > 4$  and obtained  $R=0.49(1)$. This value  to be compared with respective value
for $3d$ $SU(2)$. The monopole density $\rho_3$ in this theory was computed in  \cite{Bornyakov:1992se} : $ \rho_3 = 0.0078(2) g_3^6$.
Let us note that this value was obtained with the overrelaxation gauge fixing algorithm which, as is well known from $4D$ studies
\cite{Bali:1996dm}, gives overestimated value for the density. Using the value for $3d$ string tension $\sigma_3$ from
\cite{Teper:1993gm} $\sqrt{\sigma_3}=0.3353 g_3^2$ we obtain for the ratio
 $\rho_3^{1/3}/\sigma_3^{1/2} = 0.59(1)$. This is in good agreement with our value for $R$ if we take into account that the value for the $3d$
density is overestimated as was discussed above.

Thus in this section we have demonstrated that the density of monopoles changes with temperature in good agreement with predictions of the
dimensional reduction and being  measured in units of $\sigma_s^{3/2}$ is not far from  its  infinite temperature limit even at $T/T_c=1.5$.


\section{Study of the interaction between monopoles.}

\subsection{The details of the calculation.}

In this section we study the interaction between wrapped monopoles, through the
measurement of the spatial correlation function $g(r)=\langle \rho(0) \rho(r) \rangle / \rho^2$
for the thermal monopole-(anti)monopole pairs. On the lattice this correlation function
can be measured as the ratio of the number of the (anti)monopole located at distances $(r, r+dr)$
from a given monopole and the number (anti)monopoles in the volume $4 \pi r^2 dr$ in
a completely homogeneous system
\beq
g(r) = \frac 1 {\rho} \frac  {dN(r)} {4 \pi r^2 dr},
\label{eq:correl}
\eeq
where $dN(r)$ is the number of particles in a spherical shell $(r, r+dr)$, $\rho$ is the
average density of monopoles.  In order to take into account discretization
effects we calculate the volume $4 \pi r^2 dr$ as a number of three dimensional cubes
on the lattice which are located in the spherical shell $(r, r+dr)$. Note that 
at large distances the densities $\rho(0), \rho(r)$ become uncorrelated and 
$g(r)_{r \to \infty} \to 1$

\begin{figure}[tb]
\centering
\includegraphics[width=6.0cm,angle=270]{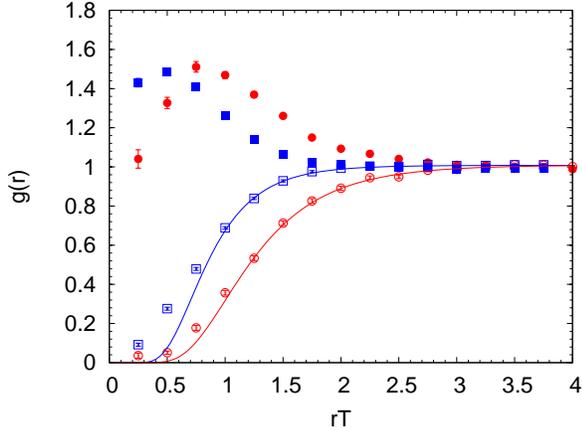}
\caption{The correlation functions $g(r)$ for $T/T_c=1.5$ (squares) and
4.8 (circles) for the monopole-monopole (empty symbols) and monopole-antimonopole
(full symbols) cases.}
\label{fig:cor}
\end{figure}

We have measured the correlation functions at the following temperatures
$T/T_c=1.5, 2.0, 3.0, 4.8, 6.9$ where our statistics was large enough.
The plots of the correlation functions at temperatures $T/T_c=1.5, 4.8$
are shown in Figure {\ref{fig:cor}}
It should be noted that the plots of the correlation functions shown
in Figure {\ref{fig:cor}} are in agreement with the results of paper
\cite{D'Alessandro:2007su} where the same correlation functions were studied.

\begin{figure}[tb]
\centering
\includegraphics[width=6.0cm,angle=270]{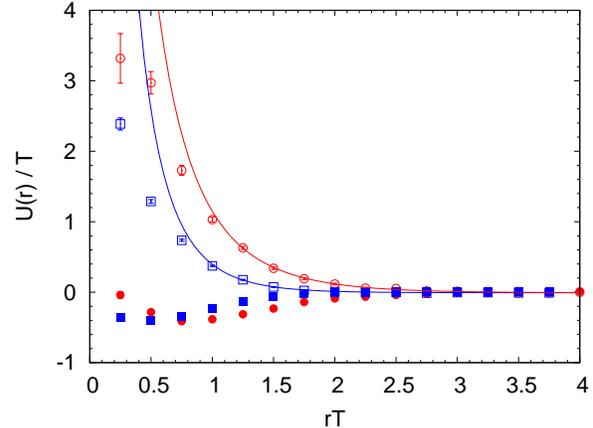}
\caption{The potentials $U(r)$ for $T/T_c=1.5$ and 4.8. The symbols are the same as in
Fig.~\ref{fig:cor}.}
\label{fig:poten}
\end{figure}

The correlation functions $g(r)$ contain information about interaction
properties of monopoles in QGP.  In order to get an idea about 
monopole-monopole and monopole-antimonopole interaction potentials
we use the  ansatz based on Boltzmann distribution
\beq
g(r)=\exp { \biggl ( - \frac {U(r)} {T}  \biggr ) },
\eeq
where $U(r)$ is the interaction potential of monopoles in QGP. Having the correlation
function one can extract the interaction potential $U(r)$. The interaction
potentials of monopole-(anti)monopole pairs for the lowest
temperature $T/T_c=1.5$ and for the temperature $T/T_c=4.8$
are shown in Figure \ref{fig:poten}. The potentials of the monopole-(anti)monopole pairs
interaction at other temperatures studied in this paper are similar to that
at temperatures $T/T_c=1.5, 4.8$. 

\subsection{The size of monopole.}

 From Figure \ref{fig:poten} it is seen that monopole-monopole repel from each other
for all values of $r$. The interaction potential of monopole-antimonopole has rather complicated
structure. At large distance the monopole-antimonopole attracts. However, the potential
reaches minimum and after the minimum it rises, becoming repulsive.

\begin{figure}[tb]
\centering
\includegraphics[width=6.0cm,angle=270]{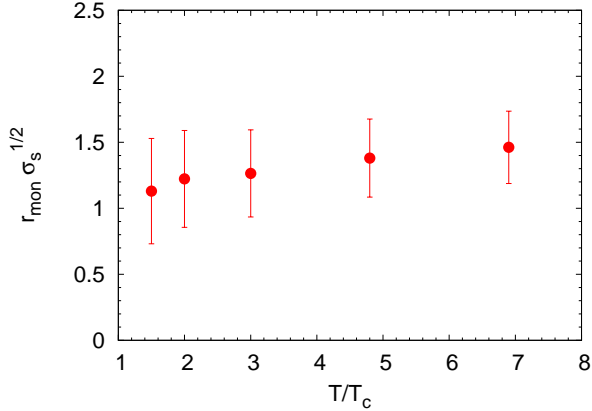}
\caption{The product $r_{mon} \sqrt{\sigma_s}$ as a function of temperature.}
\label{fig:size}
\end{figure}

Qualitative features of the monopole-antimonopole potential are
similar to that in dilute one atomic gas.
In the later case the interaction between molecules is attractive at large distances,
reaches minimum at some point and becomes repulsive at distances of order of the size of
molecule. From this analogue one can assume that  monopoles
have some size and interactions becomes repulsive at distances of order of the monopole
size $r_{mon}$. In this paper we determine the size of the monopole as $2 r_{mon} = r_{min}$, where
$r_{min}$ is the position of the minimum of the interaction potential.
In Figure \ref{fig:size} we plot the product $r_{mon} \sqrt{\sigma_s}$ as a function
of temperature. Is is seen that within the error of the calculation the plot does not
contradict to the behaviour which is suggested by the dimensional reduction at large
temperatures. However, because of the uncertainty we cannot state that the dimensional reduction
really takes place. To summarize the result of this study:  monopole is not
a pointlike object but an object with nontrivial core with the size $\sim 0.05-0.1$ fm
depending on the temperature.

\subsection{The tails of the correlation functions.}

Now let us pay attention to the tail of the correlation functions at $r \to \infty$.
In the numerical analysis we are going to fit the tail of the correlation
function as it is fitted in the
electromagnetic plasma
\begin{equation}
g(r)= \exp {\biggl (
- \frac {\alpha} {T r} e^{ - m_D r }
\biggr)},
\label{eq:fitpoten}
\end{equation}
where $m_D$ is the Debye screening mass, $\alpha$ is the monopole coupling constant. Now two comments are in order
\begin{itemize}
\item In electromagnetic plasma the constant $\alpha$ is the square of  ion charge. 
However, in QGP the interaction is much more involved. In particular, 
the mechanism of Debye screening is not the same as it is in electromagnetic plasma. 
What is more important: in view of our recent finding about dyon nature 
of monopoles \cite{Bornyakov:2011th}, one can state that monopoles interact both by chromoelectric and 
chromomagnetic charges. These facts mean that the constant $\alpha$ cannot be considered as a 
square of chromomagnetic charge. 
\item As was noted above  correlation functions (\ref{eq:correl})
are normalized as $g(r)_{r \to \infty} \to 1$. However, in lattice 
simulation the correlation functions at large distances differ 
from unity by amount $\sim  0.01$. Note the same effect 
was observed in paper \cite{Liao:2008jg}. Analysing available data we have come to 
the conclusion that this is finite volume effect. In order to take 
into account this effect we fit our data by the function
\begin{equation}
g(r)= \exp {\biggl (
- \frac {\alpha} {T r} e^{ - m_D r } 
\biggr)} + c,
\label{eq:fitpoten1}
\end{equation}
\end{itemize}

 It is natural to expect that in the limit $r \to \infty$ monopole-monopole
and monopole-antimonopole correlators have equal $M_D$ and up to 
the sign equal $\alpha$. Unfortunately we cannot check this since our lattices
have finite sizes. It is also important that the monopole-antimonopole correlator
has rather nontrivial behaviour changing from attraction to repultion at some point $r_{min}$. 
Evidently to get reliable estimation of the $M_D$ and $\alpha$ at the tail 
of the correlator one must consider the distances $r \gg r_{min}$ what is also 
impossible due to the finite lattice sizes. For this reason we determine the 
parameters $M_D$ and $\alpha$ from the monopole-monopole correlator.

\begin{figure}[tb]
\hspace*{-1.cm}
\centering
\includegraphics[width=9.0cm,angle=0]{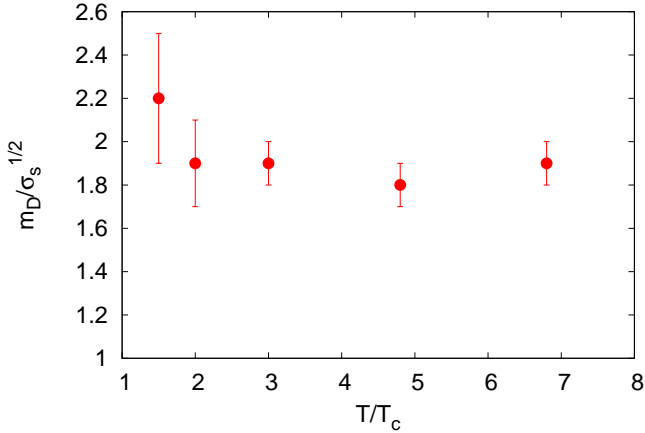}
\caption{The ratio of $M_D$ to $\sqrt{\sigma_s}$ as a function of temperature.
}
\label{fig:mass_sig}
\end{figure}

 In Table \ref{tab:massa}
we present the result of the fit for the monopole-monopole potential.
It is seen that the fit is rather good.  In Figure \ref{fig:mass_sig}
we plot the $M_D/ \sqrt{ \sigma_s}$ as a function of $T/T_c$. From this plot one sees 
that within the error of the calculation the ratio $M_D/ \sqrt{ \sigma_s}$ does not
depend on the temperature what is suggested by the dimensional reduction at large
temperatures.  

The authors of paper \cite{Liao:2008jg} studied the temperature dependence 
of the constant $\alpha$. They found that the constant $\alpha$ rises with 
temperature. Our results confirm the 
fact of rising the $\alpha$. It should also be noted that
 the values of the constant $\alpha$ obtained in \cite{Liao:2008jg} and in this work
are a little bit different but in a reasonable agreement with each other.

The authors of paper \cite{Chernodub:2004qp} found the
value $M_D/ \sqrt{ \sigma_s}=1.77$. As it is seen from Figure \ref{fig:mass_sig}
our result for this ratio is $M_D/ \sqrt{ \sigma_s} \simeq 2$ what is in good
agreement with paper \cite{Chernodub:2004qp}, taking into the account the fact
that actually the definitions of the $M_D$ in the both papers are  different.
In paper \cite{D'Alessandro:2007su}
the Debye mass was estimated  $1/ M_D \sim 0.1$ fm. Our result for the $1/M_D \sim 0.1-0.2$ fm
depending on the temperature, what is also in reasonable agreement with the estimation
of paper \cite{D'Alessandro:2007su}.



\begin{table}[ht]
\begin{center}
\begin{tabular}{|c|c|c|c|c|c|} \hline
 $T/T_c$ & $M_D/T$ & $\alpha$ & $\chi^2/N_{dof}$ \\ \hline
  1.5   & $2.2 \pm 0.3$ & $3.6 \pm 1.3$ & 1.2   \\
  2.0  & $1.9 \pm 0.2$ & $4.0 \pm 1.0$  & 1.5  \\
  3.0  & $1.7 \pm 0.1$ & $4.3 \pm 0.5$  & 0.5    \\
  4.8   & $1.4 \pm 0.1$ & $4.7 \pm 0.7$  & 1.3    \\
  6.9   & $1.4 \pm 0.1$ &  $6.6 \pm 1.5$ & 0.6   \\ \hline
\end{tabular}
\end{center}
\caption{The masses $M_D$ and coupling constants $\alpha$ obtained by fitting of  the
correlators to eq.(\ref{eq:fitpoten}).
}
\label{tab:massa}
\medskip \noindent
\end{table}

\section{The monopole mass.}

There are a lot of definitions of the monopole mass (see \cite{D'Alessandro:2010xg}). 
In this paper we are going to apply the definition which is based on the formalism 
of functional integration. Consider the trajectory of free nonrelativistic 
particle of mass $m$. The spatial fluctuation of the trajectory of such particle
\beq
\Delta { r}^2=T \int_0^{1/T} dt \langle ( \vec r(t) - \vec r(0) )^2 \rangle, 
\label{Delta_r}
\eeq
is directly connected to the particle mass 
\beq
m=\frac 1 {2 T \Delta { r}^2}.
\label{mass_m}
\eeq
The lattice version of the equation (\ref{Delta_r}) can be written as 
\beq
\Delta { r}^2=\frac 1 L \sum_{i=1}^L \langle ( \vec r_i - \vec r_1 )^2 \rangle.
\label{Delta_r1}
\eeq
Where the sum is taken along the monopole trajectory, $L$ is the total length of the 
trajectory. 

With equations (\ref{mass_m}), (\ref{Delta_r1}) we get the result presented in Table \ref{tab:massm}.
Kinetic energy of monopole is $\sim T$. Monopole mass is $\sim 2.7 T$ at $T/T_c=1.5$ reaching $\sim 8.2 T$
at $T/T_c=6.9$ what legitimates nonrelativistic approximation. 
It should be noted that the authors of paper \cite{D'Alessandro:2010xg} also calculated the mass 
(\ref{mass_m}). Their result approximately $30 \%$ larger than ours.

\begin{table}[ht]
\begin{center}
\begin{tabular}{|c|c|c|c|} \hline
 $T/T_c$ & $  a^{-2} \Delta { r}^2 $ & $m/T$  \\ \hline
  1.5   & $2.97 \pm 0.01$ & $2.69 \pm 0.01$ \\
  2.0  & $2.12 \pm 0.02$  &  $3.77 \pm 0.04$\\
  3.0  & $1.53 \pm 0.01$  & $5.23  \pm 0.03$  \\
  4.8   & $1.152 \pm 0.003$ & $6.94 \pm 0.02 $   \\
  6.9   & $0.975 \pm 0.003$ & $8.21 \pm 0.03$ \\ \hline
\end{tabular}
\end{center}
\caption{The monopole masses $m$ and spatial fluctuations at different temperatures.}
\label{tab:massm}
\medskip \noindent
\end{table}

\begin{figure}[tb]
\centering
\includegraphics[width=6.0cm,angle=270]{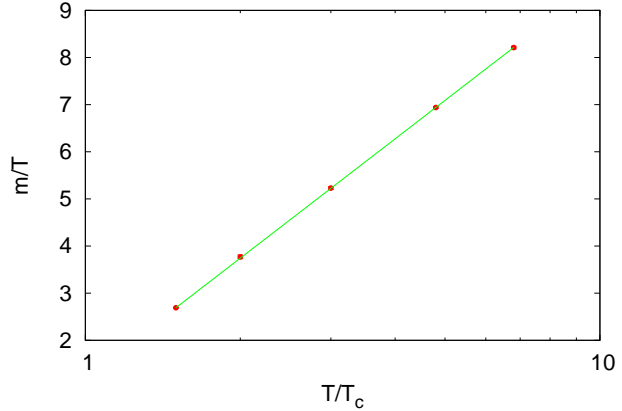}
\caption{The monopole mass $m/T$ as a function of temperature $T/T_c$ in a logarithmic scale.}
\label{fig:mass_mon}
\end{figure}

It should be noted that the larger the temperature the more static monopole trajectories. 
What means that contrary to the behaviour of the screening mass $M_D/T$
the monopole mass $m/T$ increases with temperature. In Fig. \ref{fig:mass_mon} we plot 
the monopole mass as a function of temperature $T/T_c$ in logarithmic scale. 
From this figure one sees that all points beautifully\footnote{Note that the uncertenties of the data shown in Table \ref{tab:massm} are very small.} lie on the line
\beq
\frac m T = b \log { \biggl (\frac {T} {\Lambda_m} \biggr )},
\eeq
with $b=3.653(6), \Lambda_m/T_c=0.718(2),~~ \chi^2/dof \simeq0.2$. This means that $m/T \sim 1/g^2(T)$ or $\Delta x^2 T^2 \sim g^2(T)$.

\section{The continuum limit.}

  To check the continuum limit of our results we have generated 1000 configurations of lattice 8$\times$48$^3$ with $\beta=2.635$. 
The temperature of these configurations is $T/T_c=1.5$ and it is equal to 
the temperature of the configurations at lattice 4$\times$36$^3$ with $\beta=2.43$
but the lattice spacing is two times smaller. 

  First we note that the monopole density calculated  at lattices with $a=0.1$ fm and $a=0.05$ fm
differs from each other by $\sim 5 \%$. What can be considered as a good continuum scaling and 
that even at lattice with $a=0.1$ fm we are not far from the continuum limit.

\begin{figure}[tb]
\centering
\includegraphics[width=6.0cm,angle=270]{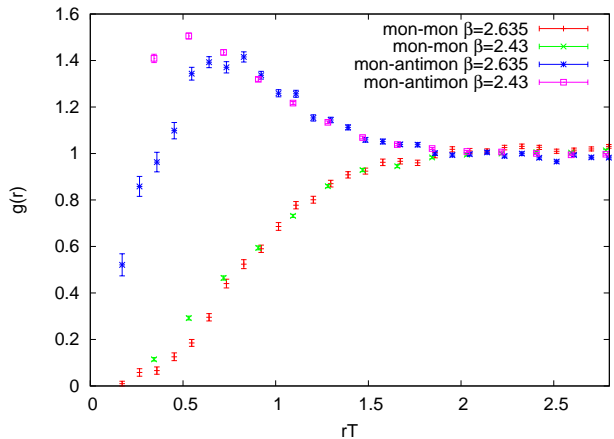}
\caption{ Monopole-monopole and  monopole-antimonopole correlators at temperature $T/T_c=1.5$ 
for lattices 8$\times$48$^3$ with $a=0.1$ fm and $\beta=2.635$ with $a=0.05$ fm.}
\label{fig:cor_cont}
\end{figure}

Further let us consider monopole correlators. In Fig. \ref{fig:cor_cont} we plot the correlators 
at different lattice spacings. From this plot one sees that both monopole-monopole and monopole-antimonopole
correlators are in good agreement with each other at distances $rT \geq 0.7$. At distances $rT \leq 0.7$
the correlators differ from each other. However, this difference can be attributed to the fact 
that the lattice spacing $a=0.1$ fm is too large and the lattice to rude to probe internal monopole structure.
At the same time the lattice spacing $a=0.05$ fm is small enough to do this. Note that the domain of repultion 
of the monopole-antimonopole pair is clearly seen on this plot. 

In section IV the size of monopole was defined as $2 r_{mon}=r_0$, where $r_0$ is the position 
of the maximum in the monopole-antimonopole correlator. The position of the maximum shifted to the right
but not greater than by $20 \%$. If the maximum were lattice artefact for dimensional arguments 
its position would shift to
the left. From this we conclude that monopole has some physical size which has good scaling behaviour. 

The tail parameters $\alpha, M_D/T$ (\ref{eq:fitpoten1}) and monopole mass $m/T$ (\ref{mass_m}) at lattice 
with $a=0.05$ fm have the following values
\beq
\alpha=3.0 \pm 0.8, \quad \frac {M_D} T = 1.9 \pm 0.2, \quad \frac m T = 3.96 \pm 0.2.
\label{res_cont1}
\eeq
We see that taking into the account uncertainties of the calculation the results for the 
$\alpha, M_D/T$ are in agreement with Table \ref{tab:massa}. As to the monopole mass $m/T$ its 
value is by $30 \%$ larger than the result presented in Table \ref{tab:massm}. Note that 
Table \ref{tab:massm} was obtained at lattices with $N_t=4$, but  results Table \ref{res_cont1}
were obtained at lattice with $N_t=8$. Evidently, the thermal monopole trajectory parameters are very 
sensitive to the $N_t$. So, this could be the reason of the deviation
of the both results.

\section{Conclusion.}

In this paper the properties of the thermal abelian monopoles are studied in the deconfinement phase of
the $SU(2)$ gluodynamics. To remove effects of Gribov copies the simulated annealing algorithm is
applied for fixing the maximally abelian gauge.

We calculated the density of thermal abelian monopoles
at different temperatures. It was found that at temperatures $T> 3 T_c$ our data can be very well
fitted by the function which is motivated by dimensional reduction prediction $\rho \sim (g^2(T) T)^3$.

I addition to the monopole density in this paper we studied the monopole-(anti)monopole spatial correlation
functions. We calculated the correlation functions at the temperatures $T/T_c=1.5, 2.0, 3.0, 4.8, 6.9$.
This result allowed us to determine the potentials of monopole-monopole and monopole-antimonopole 
interactions. As one can expect monopoles repel from each other.
The tails of the interaction potential can be well fitted by the screened Coulumb potential
\beq
U(r)=\frac {\alpha} r e^{-M_D r}.
\eeq
We determined the dependence of the parameters $\alpha, M_D$ on temperature. 

The interaction potential of monopole-antimonopole has rather complicated
structure. At large distances it is attractive. However, the potential
has a minimum at some distance and after the minimum it rises, becoming repulsive. The position
of the minimum can be considered as a double monopole size. In this way we estimated
the size of monopole which turned out to be $0.05-0.1$ fm depending on the temperature.

We have also calculated the mass of monopole and the dependence of the mass on temperature. 

The last point considered in this paper is the continuum limit of the results obtained in this paper. 
The calculation was done at lattice 8$\times$48$^3$  with two times smaller lattice spacing $a=0.05$ fm in 
comparison to the lattice 4$\times$36$^3$ (see Table \ref{tab:statistics}). 
All results except the monopole mass have good scaling behaviour. The monopole mass is increased by 30\%. 
To undestand the origin of this increase further study is needed.

 \subsection*{Acknowledgments}
We would like to express our gratitude to M.I. Polikarpov and V.I. Zakharov for very useful and illuminating discussions.
This investigation has been supported by the Federal Special-Purpose
Programme 'Cadres' of the Russian Ministry of Science and Education,
by the grant for scientific schools NSh-6260.2010.2 and by grant
RFBR 11-02-01227-a. VVB is supported by grant RFBR 10-02-00061-a and RFBR 11-02-00015-a. VGB is
supported by grant RFBR 09-02-00338-a.

\bibliographystyle{apsrev}
\bibliography{citations_monop}

\end{document}